\documentstyle[twocolumn,prl,aps]{revtex}
\large

\begin{document}
\draft
\twocolumn[
\hsize\textwidth\columnwidth\hsize\csname @twocolumnfalse\endcsname

\draft
\title{
     Connected network of minima as a model glass: long time dynamics
      }
\author{
	L.~Angelani$^{1}$,
        G.~Parisi$^{2}$,
         G.~Ruocco$^{1}$ and 
	G.~Viliani$^{3}$
       }
\address{
	 $^{1}$ 
	 Universit\'a di L'Aquila and Istituto Nazionale di Fisica 
	 della Materia, I-67100, L'Aquila, Italy. \\
	 $^{2}$ 
	 Universit\'a di Roma {\rm La Sapienza} and Istituto Nazionale di 
         Fisica Nucleare, I-00185, Roma, Italy. \\
	 $^{3}$ 
	 Universit\'a di Trento and Istituto Nazionale di Fisica
	 della Materia, I-38050, Povo, Trento, Italy.\\
	}
\date{\today}
\maketitle
\begin{abstract}
A simple model to investigate the long time dynamics of glass-formers is 
presented and applied to study a Lennard-Jones system in supercooled and 
glassy phases.
According to our model, the point representing the system in the 
configurational phase space performs harmonic vibrations around 
(and  activated jumps between) minima pertaining to a connected network. 
Exploiting the model, in agreement with the experimental results, 
we find evidence for: {\it i)} stretched relaxational dynamics; 
{\it ii)} a strong $T$-dependence of the stretching parameter; 
{\it iii)} breakdown of the Stokes-Einstein law.
\end{abstract}
\pacs{PACS Numbers : 61.20.Lc, 64.70.Pf, 82.20.Wt}
]
In recent years many efforts were devoted to the understanding of the 
phase space landscape in supercooled liquids and structural glasses, and, in
particular, to the identification of those landscape details that are 
responsible for the structural arrest taking place at the glass transition 
temperature, $T_g$ \cite{landscape}. It has been 
recently speculated \cite{equiv} that the free energy landscape 
of structural glasses is similar to that of some generalized spin glasses 
models, where it was shown that exists a dynamical temperature
$T_{D}$ (which is well defined in mean field 
approximation and becomes a crossover region in real systems) below
which 
the dynamics is dominated by long time activated processes consisting of 
jumps among different free energy minima. The parallel between  $T_D$ 
and  the critical temperaure, $T_C$, of the mode cupling theory (MCT)
\cite{MCT} is natural. If verified, this parallel could lead to a 
microscopic description of the dynamic slowing down, predicted by MCT, 
in term of free energy landscape. Approaching $T_C$ (or $T_D$), however, 
the presence of this very slow dynamics makes the numerical investigation 
of structural glasses very hard. To our knowledge,  only few attempts were 
made in this direction \cite{slowdyn}.

In this letter we introduce a new method to study the slow dynamics in 
glasses and in deeply supercooled liquids; at variance with  the usual 
Molecular Dynamics (MD) simulation we describe the dynamics of the
system as relaxations taking place in a connected network of
potential energy minima. The jumps among minima are described by an
appropiate master equation, and, in this way, we can investigate the 
long time behaviour of a glass in short simulation times as the solution
of the master equation is an eigenvalue problem. 
The characteristics and connectivity of the minima, and 
other energy-landscape properties entering the determination of the 
transition probabilities, are inferred from the MD investigation 
of a small system (one component Lennard-Jones in the present case).

The physical quantities (total energy, pressure, transport coefficent) 
obtained from the model agree with those derived from MD up to 
temperature above the melting point, supporting the jump model even in the
liquid phase of Lennard-Jones fluids.
In the low temperaure region we found evidence for:
{\it i)}  stretched behaviour of the relaxation process; 
{\it ii)}  temperature dependence of the stretching exponent, $\beta_K$,
which changes from $\approx 1$ at high $T$ down to $\approx 0.3$ at low 
$T$; and {\it iii)}  breakdown of the Stokes-Einstein 
relation. All these results are in agreement with the experimental 
findings in real "fragile" \cite{angell} glasses.
 
In a glass, the atoms are
(almost) frozen in some (meta)stable positions. The short-time dynamics is 
dominated by small vibrations around the stable position. This dynamics 
can be described within the harmonic approximation, and all the relevant
information is obtained by diagonalizing the dynamical matrix. 
At long time, collective jumps among different stable positions involving
many atoms 
become possible and are controlled by a master equation. These long time
relaxations are only apparently in
contraddiction with 
the {\it harmonic} vibrational dynamics within the local minima; 
indeed, it has been recently shown that the relevant classical path for
relaxation between adjacent minima is practically decoupled from the other
degrees of freedom, which are almost harmonic \cite{DEMIC}.
The transition rates, in turn, are determined by
minima energies, barrier heights and other topological properties.

In order to set up the connected netwok of minima and to determine the 
transition rates we need the topology of the
multidimensional potential energy hypersurface of the system. 
To this end, we numerically analyse the Potential Energy Landscape 
(PEL) of small ($N=11 \div 29$ atoms) Lennard-Jones systems whit periodic
boundary conditions. The small size of the system allows us an exhaustive
investigation of  the landscape but, at the same time, exhibits 
complex enough features and behaviour to capture the physics of the system.
The atoms interact via the $6-12$ Lennard-Jones potential
 $V_{LJ}(r)=4\epsilon[(\frac{\sigma}{r})^{12} - (\frac{\sigma}{r})^{6} ]$,
whit $\epsilon/K_{B} = 125.2$ K ($K_{B}$ Boltzmann's constant) and 
$\sigma = 0.3405$ nm, appropriate for Argon.  The simulated density is 
$\rho = 42$ mol/dm$^{3}$. Due to the small dimensions of the system, we
choose the multi-image method, 
namely each particle interacts with many images of each other (in practice, 
due to the cutoff distance that we impose on the pair potential, 
$r_{cut} = 2.6\ \sigma$, there are at most $3$ interacting images).  

The first step is to search for the potential energy minima.
We perform a modified steepest descent method  procedure \cite{DEMIC},
starting from high temperature MD configurations to find 
the inherent configurations corresponding to local minima which often 
result to be crystalline-like. In order to
establish whether a  minimum corresponds to a glassy structure, we use
the static stucture factor
$S(\vec{q})=N^{-1}\mid \sum_{j}e^{i \vec{q}\cdot\vec{r_{j}}}\mid^{2}$.
For a pure crystalline configuration of $N$ particles $S(q)$ is 
made up by 'Bragg' peaks and its value at the peaks is $S_{max}=N$,
whereas for a glass one usually finds $S_{max} \approx 2 \div 3$. In small
size samples there are obviously intermediate situations and for a minimum
to be 'glassy' we adopt the criterion $S_{max}\leq N/2$.

As a second step, for each pair of minima, $a$ and $b$, with 
energy $E_a$ and $E_b$ respectively,  we first determine the mutual distance
$d_{a,b} = min (\vert \underline{r}_{a} - \underline{r}_{b}\vert)$, 
where $\underline{r}_{a}$ is the position vector of the minima in the $3N$
dimension configurational space, and $min$ indicates the minimization 
with respect to the all symmetry operations: continuous translations,
permutations of particles and the 48 symmetry operations of the cubic 
group. Then we analyse the potential energy profile experienced by the 
system in travelling from one minimum to another and determine the
potential energy barrier. Among the different paths joining $a$ and $b$, 
we assume \cite{DEMIC} that the system follows that with the least action.
The action integral is defined as
$S(\ell) = \int_{\ell} ds \sqrt{V(\underline{r}(s)) - V_{o}}$,
where $\ell$ indicates a generic path, $s$ the curvilinear coordinate, 
and $V_{o}= \min \{E_a,E_b\}$. The minimization of $S(\ell)$ is performed
by dividing the path in $n=16$ intervals  and minimizing the action function
with respect to the
extrema of the $n$ segments constrained to move in iperplanes perpendicular
to the straigh path.  The highest energy value, $V_{ab}$, along the
least action path (LAP) determines the saddle point of the path. Not all 
the pairs of minima are directly connected, since, sometimes, the LAP
joining them crosses a third minimum. Therefore, there is a non-trivially
connected network of minima.
Next we mesure the curvature, defined as the determinant of the Hessian of
potential energy function, in each minimum $a$ ($\det \{V''_a\}$),
and $a-b$ saddle point ($\det \{V''_{ab}\}$). Also important is
the absolute value of the negative curvature on the saddle point, 
$\tilde\omega_{ab}$.

In order to give a full statistical description of the PEL, we
study the distributions, $P(x)$, of the relevant quantities $x$ 
(here $x$ represents $E_a$, $\Delta E_{ab}=E_a-E_b$, $d_{ab}$, $V_{ab}$, 
$\det \{V''_a\}$, $\det \{V''_{ab}\}$, or $\tilde\omega_{ab}$)
and their cross correlations, $P(x_1,x_2)$. Cross-correlations among the
measured quantities are observed. The most evident is 
a linear correlation in double $\log$ scale between the distance and 
the barriers' height along the LAP between two minima. 
A rather weak correlation is also observed between the energy and 
curvature of extrema points of PEL.

The model we introduce is a connected network of potential energy minima 
with a jump-dynamics described by an appropriate master equation:
\begin{equation}
\dot p_{a}(t) = \Sigma_{c} W_{ac} \ p_{c}(t) \ ,
\end{equation}
where $p_{a}(t)$ is the probability that the system is in minimum $a$ 
at time $t$ (actually $p_{a}(t) \equiv p_a(t|b)$,
indicating that at $t=0$ the system was in minimum $b$) 
and the non-diagonal elements of the transition matrix, the transition rates
$W_{ab}$, are determined from the energetic and topological properties 
of the PEL. In order to satisfy the equilibrium condition, 
$p_a^o=p_{a}(t\rightarrow\infty) \propto 
(\det V''_{a})^{-1/2} \ e^{-\beta E_{a}}$, with $\beta = (K_{B} T)^{-1}$,
$W_{ab}$ must obey the detailed balance
$W_{ab} \ p_{b}^o = W_{ba} \ p_{a}^o$.
Following \cite{RISK} we make the ansatz:
\begin{equation}
W_{ab} = \frac{\tilde\omega_{ab}^{2}}{\gamma}
\ \left(\frac{\det V''_{b}}
  {\mid \det V''_{ab} \mid}  \right)^{1/2}
  \ e^{-\beta (E_{ab}-E_{b})}\ ,
\end{equation}
where $\gamma$ is a friction constant which actually determines the
time scale.
This choice of the transition matrix is based on the 
approximation  of the problem of escape from a metastable state as
a Markovian Brownian  multidimensional motion in the overdamped
friction regime \cite{RISK}.

To set up our model minima-network we proceed in the following way. 
Having fixed the number of minima ($M$=400 in the present case),
we extract the energy of these minima and their curvatures
from the previously found bivariate distribution.
For each minimum we randomly extract $20$ minima connected to it
and define a connection matrix $c_{ab}$ that contains the number
of steps required to go from $a$ to $b$.
We then define the distance $d_{ab}$ as $c_{ab}$ times the value 
extracted from the distribution of the distances $P(d_{ab})$,
and from these the energies of saddle points.  The further statistical 
features of saddle points (curvatures) and minima (trasverse component 
of the microscopic stress tensor, see below) are determined from 
bivariate (correlation curvature-saddle point energy) and simple 
extractions, respectively.

To check the reliability of the model, we first concentrate on 
the static properties. Following our model, the partition function is 
approximated by a sum over the minima and the harmonic vibrations 
around them:
\begin{equation}
 \label{parfun}
{\cal Z}(\beta) = \beta^{-3N/2} \Sigma_{a}(\det V''_{a})^{-1/2}  
  \ e^{-\beta E_{a}}.
\end{equation}

In fig. 1 we show the potential energy of a LJ system with 
$N=29$ particles as obtained through MD and as calculated
from (\ref{parfun}) by taking into account either all minima (dotted line),
or only the glassy ones (full line). The MD data are obtained
progressively heating the glass ($\circ$) up to the liquid phase,
and then cooling the system slowly ($\bullet$), to obtain
crystallization. We observe a quantitative agreement between the 
MD data and the model up to $T \approx 150$ K, a temperature well 
above the melting point ($T_m \approx 80$ K). At higher
temperature the simple local-vibration/collective-jumps model
fails.

As for the dynamical properties, the solution of the master equation 
is found by numerical knowledge of eigenvalues and eigenvectors of 
the transition matrix $\lambda(n)$ and $v_{a}(n)$, with $n=1...M$.
In particular:
\begin{equation}
p_{a}(t|b) = \sum_{n} \frac{v_{a}(n)\ v_{b}(n)}{p_{b}^{o}}
\ e^{\lambda(n) t} \ .
\end{equation}
It is then possible to determine the statistical equilibrium average of 
a generic observable $O(t)$ from the knowledge of its value, $O_{ab}$,
calculated at the minima $a$ and $b$:
\begin{equation}
\label{valormedi}
<O(t)> = \sum_{b} p_{b}^{o} \sum_{a} O_{ab} \ p_{a}(t|b).
\end{equation}

We consider three observables: 
the mass diffusion coefficient $D$, the shear viscosity $\eta$,
and the structural relaxation time $\tau$. In the first case
$O_{ab} = \mid \underline{r}_{a}-\underline{r}_{b} \mid^{2}$, and
$D= \lim_{t\rightarrow\infty} <O(t)>/6t$.
For the two other cases, we first determine the correlation function, 
of the off-diagonal elements of the stress tensor, 
$C(t) = <\sigma^{zx}(0)\sigma^{zx}(t)>$, with \cite{nota1}:
\begin{equation}
\sigma^{zx} = - \sum_{i>j} \frac{z_{ij}x_{ij}}{r_{ij}} \ V'(r_{ij}) \ .
\end{equation}
In the notation of eq. (\ref{valormedi}), 
$O_{ab}= \sigma_a^{zx} \sigma_b^{zx}$,
(where $\sigma_a$ is the value of the stress tensor calculated
at minumum $a$).
Then the shear viscosity is calculated as
\begin{equation}
\label{sv}
\eta = (K_B TV)^{-1} \int_{0}^{\infty} dt \ C(t),
\end{equation}
and the relaxation time $\tau$ is derived from a fit of $C(t)$ to
a stretched exponential decay $C(t)=C(0) \exp{-(t/\tau)^{\beta_K}}$.

We report the values obtained for $M=400$ and avereged over $50$ 
different extraction of the network parameters.
In fig. 2 we show the normalized correlation functions $C(t)/C(0)$ 
calculated at different temperatures together with their best fits. 
In the inset the $T$-dependence of the stretching parameter $\beta_K$ 
is also reported. We remind that our model reproduces only the 
structural ($\alpha$) relaxation processes usually explained in term of
{\it intra-basins} transitions, at variance with other fast relaxation 
processes taking place {\it inside} the basins. 
This explain the presence of only one step relaxation in $C(t)$.
We also observe that: {\it i)} the relaxation dynamics is
well represented by a stretched exponential decay; and {\it ii)}
the stretching parameter $\beta_K$ is strongly temperature
dependent, implying a violation of the time-temperature 
superposition principle \cite{VtT}. Moreover $\beta_K$ decreases from 1 at
high $T$ (Debye relaxation) down to $\approx$ 0.35 at low
$T$, a value consistent with experimental findings \cite{exp_beta}
and theoretical prediction \cite{theo_beta} in fragile glass-formers.

In fig. 3a we show the shear viscosity and the relaxation time $\tau$
versus inverse temperature. They are almost proportional to each other,
strongly increasing in a small temperature range ($150-20$ K). 
However, their temperature behaviour is well represented by an Arrhenius
law,
and does not show the dramatic increase expected for fragile glass-formers.
Whether this unexpected behaviour {\it i)} has to be ascribed to a failure of
our
model, or {\it ii)} is a genuine behaviour of LJ liquids at constant density,
is still unknown.
At those temperature where the direct MD calculation of $\eta$ is affordable
($\circ$ in fig. 3a) we found a good agreement between MD's and model's results,
supporting the hypothesis {\it ii)}.

In fig. 3b we report the inverse diffusion coefficient, $D^{-1}$,
versus the ratio $\eta/T$. The Stokes-Einstein (SE) relation would 
predict direct proportionality, i.~e. $D \propto T/\eta$. 
The full line (slope 1) indicates that, at high $T$, the SE
relation asimptotically holds. Upon decreasing $T$, the slope $\xi$ decrease
towards $\xi \approx 0.28$, indicating a breakdown of the SE relation,
as observed in different experiments \cite{BSE}.
In particular, the crossover between the two regimes occours in the same
temperature region where $\beta_K$ deviates from $1$. 
It is tempting to note that the crossover position, at $\eta/T \approx 0.1$
Poise/K,
and the fractional exponent at low $T$, $\xi \approx 0.28$, are in fairly good
agreement with the experimental results in the fragile glass former 
{\it o}-terphenyl \cite{FDSE,note2}.

In conclusion, we presented a simplified model, based on a vibrational local
dynamics and on collective jumps among minima, that well describes the
structural
relaxation features of supercooled liquids and glasses. Exploting this model
we are 
able to investigate the long time (low temperatures) dynamics. We recover,
in  
the simple LJ system, some important features of real glass-former, 
in particular: {\it a)} stretching of the relaxational dynamcs; {\it b)}
failure
of the time-temperature superposition principle ($T$-dependence of $\beta_K$);
and 
{\it c)} breakdown of the Stokes-Einstein relation.

We wish to thank B.~Coluzzi, G.~Monaco and F.~Sciortino for useful discussions,
and D.~Leporini who called our attention to the fractional SE relation.

{\footnotesize{
\begin{center}
{\bf CAPTIONS}
\end{center}

\begin{description}

\item  {Fig. 1 - 
Potential energy of the LJ system as determined from MD heating the glass
($\circ$) and cooling the liquid ($\bullet$), and from the present
model using all the minima (dotted line) or only the glassy minima
(full line).
}

\item  {Fig. 2 - 
Normalized autocorrelation functions of the off-diagonal elements of 
the stress tensor at the indicated temperature as determined from the
model (open symbols). The lines represent the best fits to the 
data with a stretched exponential time decay. The inset show the
$T$ dependence of the stretching parameter $\beta_K$.
}

\item  {Fig. 3 - 
a) Shear viscosity and relaxation time versus inverse temperature.
b) Inverse diffusion coefficient, $D^{-1}$, versus the ratio $\eta/T$.
According to the Stokes-Einstein relation, a linear proportionality
is expected (dashed line). The full line, with slope 0.28, is the best
fit to the low temperature data.
}
\end{description}
}
}

\begin{references}
\bibitem{nota1}
In the definition of stress tensor we have omitted the velocity 
term, not well defined in our model.
\end{references}
\end{document}